\documentclass{SciPost}

\hypersetup{
    colorlinks,
    linkcolor={red!50!black},
    citecolor={blue!50!black},
    urlcolor={blue!80!black}
}

\usepackage[bitstream-charter]{mathdesign}
\DeclareSymbolFont{usualmathcal}{OMS}{cmsy}{m}{n}
\DeclareSymbolFontAlphabet{\mathcal}{usualmathcal}

\fancypagestyle{SPstyle}{
\fancyhf{}
\lhead{\colorbox{scipostblue}{\bf \color{white} ~SciPost Physics }}
\rhead{{\bf \color{scipostdeepblue} ~Submission }}

\fancyfoot[C]{\textbf{\thepage}}
}

\begin{document}

\pagestyle{SPstyle}

\begin{center}{\Large \textbf{\color{scipostdeepblue}{
%%%%%%%%%% TODO: Write your article's title here
Quark and lepton mixing in the asymptotically safe Standard Model\\
}}}\end{center}

\begin{center}\textbf{
% Use (full) first name (+ middle name initials) + surname format.
% Separate subsequent authors by a comma, omit comma and use "and" for the last author.
% Mark the corresponding author(s) with a superscript symbol in this order
% \star, \dagger, \ddagger, \circ, \S, \P, \parallel, ...
Astrid Eichhorn\textsuperscript{1$\star$},
Zois Gyftopoulos\textsuperscript{1$\dagger$} and
Aaron Held\textsuperscript{2$\ddagger$}
}\end{center}

\begin{center}
{\bf 1} Institute for Theoretical Physics, Heidelberg University, Philosophenweg 12 and 16, 69120 Heidelberg, Germany
\\
{\bf 2} Institut de Physique Th\'{e}orique Philippe Meyer, Laboratoire de Physique de l’\'Ecole normale sup\'erieure (ENS), Universit\'e PSL, CNRS, Sorbonne Universit\'e, Universit\'e Paris Cit\'{e}, F-75005 Paris, France
%%%%%%%%%% END TODO: AFFILIATIONS
% Provide email address of corresponding author(s)
\\[\baselineskip]
$\star$ \href{mailto:email1}{\small eichhorn@thphys.uni-heidelberg.de}\,,\quad
$\dagger$ \href{mailto:email1}{\small gyftopoulos@thphys.uni-heidelberg.de}\,,\quad
$\ddagger$ \href{mailto:email2}{\small aaron.held@phys.ens.fr}
\end{center}

\section*{\color{scipostdeepblue}{Abstract}}
\textbf{\boldmath{%
The quark mixing (CKM) matrix is near-diagonal, whereas the lepton mixing (PMNS) matrix is not. We learn that both observations can generically be explained within an ultraviolet completion of the Standard Model with gravity. 
This ultraviolet completion results in approximate relations between CKM matrix elements. The accuracy with which these hold translates into the length of asymptotically safe scaling regimes. Experimental data are consistent with the theoretical expectations arising from a fixed-point cascade at transplanckian scales.
Similar results would hold for the PMNS matrix if neutrino Yukawa couplings were large. The ultraviolet complete theory, therefore, must -- and in fact can -- avoid such an outcome.
It achieves this through a dynamical mechanism that limits the size of neutrino Yukawa couplings. Thus, large neutrino mixing is intimately tied to small Dirac neutrino masses, $\sum m_{\nu} \lesssim {\mathcal{O}} (1)\, \rm eV$ and a mass gap in the Standard Model fermion masses.
}}

\vspace{\baselineskip}

\noindent\textcolor{white!90!black}{%
\fbox{\parbox{0.975\linewidth}{%
\textcolor{white!40!black}{\begin{tabular}{lr}%
  \begin{minipage}{0.6\textwidth}%
    {\small Copyright attribution to authors. \newline
    This work is a submission to SciPost Physics. \newline
    License information to appear upon publication. \newline
    Publication information to appear upon publication.}
  \end{minipage} & \begin{minipage}{0.4\textwidth}
    {\small Received Date \newline Accepted Date \newline Published Date}%
  \end{minipage}
\end{tabular}}
}}
}
%%%%%%%%%% BLOCK: Copyright information

% For convenience during refereeing we turn on line numbers:
% \linenumbers

% \vspace{10pt}
% \noindent\rule{\textwidth}{1pt}
% \tableofcontents
% \noindent\rule{\textwidth}{1pt}
% \vspace{10pt}

% Write your article contents here, starting from first \section.
% An example structure is given below.

\section{Introduction} \label{sec:intro}
% TODO: write your article here.
In the Standard Model (SM), the Cabbibo-Kobayashi-Maskawa (CKM) matrix relates the mass eigenstates of the quarks to their flavor eigenstates \cite{Cabibbo:1963yz,Kobayashi:1973fv}.
The CKM matrix is
\begin{eqnarray}
|V|_{\rm CKM}^2=\left(
\begin{matrix}
|V_{ud}|^2 & |V_{us}|^2 & |V_{ub}|^2\\
|V_{cd}|^2 & |V_{cs}|^2 & |V_{cb}|^2\\
|V_{td}|^2& |V_{ts}|^2& |V_{tb}|^2
\end{matrix}\right).
\end{eqnarray}
Unitarity requires each row and column to sum to one. The four independent elements are experimentally determined, see \cite{ParticleDataGroup:2024cfk} and \cite{Pocanic:2003pf,KLOE:2006vvm,NA482:2006vnw,CLEO:2009svp,CLEO:2009lvj,CLEO:2009vke,BaBar:2010ixw,Belle:2013isi,BaBar:2014xzf,BESIII:2016cws,BESIII:2017ylw,Markisch:2018ndu,Czarnecki:2019iwz,Hardy:2020qwl,BESIII:2021anh,Cirigliano:2022yyo,Seng:2021nar,HFLAV:2022esi,Seng:2022wcw},
\begin{eqnarray}
 \! \!\!\!\!  |V_{ud}|^2\! \!\!&=&\!\!\! 0.94936 \pm 0.00031,\; |V_{us}|^2 \!=\! 0.05063 \pm 0.00031 \nonumber\\
   \!\!\!\!\! |V_{cd}|^2\!\!\! &=&\!\!\! 0.05057 \pm 0.00031,\; |V_{cs}|^2 \!=\! 0.94768 \pm 0.00031. 
\end{eqnarray}
They approximately satisfy the relations 
\begin{eqnarray} \label{eq:IRline}
    % 1- Y = X = W = 1 - Z.
     |V_{ud}|^2+|V_{us}|^2=1, \quad |V_{cd}|^2+|V_{cs}|^2=1,\\
     \mbox{i.e., due to unitarity }\, |V_{ub}|^2 =0, \,|V_{cb}|^2=0.
\end{eqnarray}  
The same relations 
characterize a fixed line of the Renormalization Group (RG) flow, which ends in a fixed point at $|V_{ud}|^2=1=|V_{cs}|^2$, $|V_{us}|^2=0=|V_{cd}|^2$ \cite{Pendleton:1980as}. The line exists, if the top Yukawa coupling $y_t$ is the dominant Yukawa coupling, i.e., $y_t \gg y_{i\neq t}$ \cite{Alkofer:2020vtb}. 
The line attracts RG flows towards the infrared (IR). 
 We expect that the
 relations \eqref{eq:IRline} only hold up to a finite accuracy, which is determined by the rate of change of the CKM elements under the RG flow. In the vicinity of the fixed line, the rate of change is  set
 by the critical exponents, which are $\theta_{1,2,3}= -3y_t^2/(16\pi^2)$; with $\theta_4=0$ indicating a fixed line \footnote{Critical exponents for our case can be calculated as $- \frac{\partial \beta_{|V_{ij}|^2}}{\partial |V_{ij}|^2}\Big|$, evaluated at the fixed line. The beta-functions, $\beta_{|V_{ij}|^2}= \partial_t |V_{ij}|^2$ describe the RG flow as a function of the logarithmic scale derivative $\partial_t = k\, \partial_k$.}.
For the Planck-scale value $y_t \approx 0.38$, \cite{Buttazzo:2013uya} we find
$\theta_{1,2,3} \approx -
 3\cdot10^{-3}$, indicating a very slow RG flow, $ (|V_{ud}|^2+|V_{us}|^2-1)(k_{\rm IR}) =  
 (|V_{ud}|^2+|V_{us}|^2-1)(k_{\rm UV}) \cdot(k_{\rm IR}/k_{\rm UV})^{10^{-3}}$ from the ultraviolet (UV) scale $k_{\rm UV}$ to the IR scale 
 $k_{\rm IR} \approx 171\, \rm GeV$. A ten-fold change in the CKM matrix elements requires 300 orders of magnitude in scales.
Accordingly, to approach the relation~\eqref{eq:IRline}  to its realized $10^{-3}$  
 accuracy
from non-fine-tuned 
initial conditions at $k_{\rm UV}$ requires an RG flow over at least 1000
orders of magnitudes in scales, $k_{\rm UV}/k_{\rm IR}= 10^{1000}$. 

From the experimental values of the CKM matrix, we thus conclude that one of two alternatives is realized: 
\begin{itemize}
\item \textbf{Alternative (a):} UV completions of the SM do not use this mechanism to fix the CKM matrix values. The agreement between the experimental values and the IR attractive fixed line of the RG flow is pure coincidence.
\item \textbf{Alternative (b):} The SM, completed by suitable new physics, remains viable for many orders of magnitude beyond the Planck scale and $y_t$ is dominant over at least $10^3$ orders of magnitude in scales. 
\end{itemize}
We consider the second alternative more compelling.
\paragraph{Explaining the PMNS matrix --}
Any explanation of the structure of the CKM matrix begs the question why a very different structure is realized in the mixing of the leptons, described by the Pontecorvo-Maki-Nakagawa-Sakata (PMNS) matrix \cite{Pontecorvo:1957qd,Maki:1962mu}, which reads
\begin{eqnarray}
|U|_{\rm PMNS}^2=\left(
\begin{matrix}
|U_{e1}|^2 & |U_{e1}|^2 & |U_{e3}|^2\\
|U_{\mu 1}|^2 & |U_{\mu2}|^2 & |U_{\mu3}|^2\\
|U_{\tau1}|^2& |U_{\tau2}|^2& |U_{\tau3}|^2
\end{matrix}\right).
\end{eqnarray}
 For Dirac neutrinos, four elements are independent
 \footnote{Unitarity is imposed through the row-wise relations $|U_{e3}|^2=1-|U_{e2}|^2-|U_{e1}|^2$, $|U_{\mu3}|^2=1-|U_{\mu2}|^2-|U_{\mu1}|^2$, $|U_{\tau3}|^2=1-|U_{\tau1}|^2-|U_{\tau2}|^2$, $|U_{\tau3}|^2=1-|U_{\tau2}|^2-|U_{\tau3}|^2$ and similarly column-wise. The unitarity violations are experimentally constrained \cite{Antusch:2014woa}.} 
 and determined experimentally, see \cite{ParticleDataGroup:2024cfk} \cite{SNO:2011hxd,MINOS:2013utc,KamLAND:2013rgu,IceCube:2019dqi,deSalas:2020pgw,Esteban:2020cvm,Capozzi:2021fjo,DayaBay:2022orm,T2K:2023smv,Esteban:2024eli,T2K:2024wfn} as
\begin{eqnarray}
|U_{e1}|^2&=&  0.678 \pm 0.012\;\;,\;\; |U_{e1}|^2=  0.300\pm 0.012,\\
|U_{\mu1}|^2&=&  0.098 \pm 0.038\;\:,\;\; |U_{\mu2}|^2= 0.380\pm 0.051.
\end{eqnarray}
Its non-diagonal form is markedly different from the CKM matrix. A priori, this difference 
is
a problem for alternative b), because the RG flow of both matrices is determined by an equivalent set of evolution equations, which 
implies that $|U_{e3}|^2=0$ and $|U_{\mu3}|^2=0$
are IR attractive, just like  $|V_{ub}|^2=0$ and $|V_{cb}|^2=0$. To show this, we discuss an approximation to these evolution equations, see Appendix \ref{app:Bderivation} :
 Given one heavy fermion with large Yukawa coupling $y_h$ and two lighter fermions with Yukawa couplings $y_i$ and $y_j$   and corresponding Dirac masses $m_{i,j}$ with
\begin{equation}
 \frac{y_i^2+y_j^2}{y_i^2-y_j^2}= \frac{m_i^2+m_j^2}{m_i^2-m_j^2} = \frac{\Sigma m_{ij}^2}{\Delta m_{ij}^2},
\end{equation}
the RG flow of
 the matrix elements
is, see also \cite{Ohlsson:2013xva}, 
\begin{equation}
k \partial_k \left(|V_{ub(cb)}|^2\right)\approx \!\frac{3}{32\pi^2}\frac{\Sigma m_{ij}^2}{\Delta m_{ij}^2} y_h^2 \, |V_{ub(cb)}|^2+...,\label{eq:approxflowmixing}
\end{equation}
and similarly for $|U_{e3}|^2$ and $|U_{\mu 3}|^2$.
Under the RG flow towards the IR, these four matrix elements
decrease, approaching relation \eqref{eq:IRline}.
The rate of change of the matrix elements
is set by the product $\Sigma m_{ij}^2/\Delta m_{ij}^2 \,\cdot y_h^2$.  
As a result, $|V_{ub(cb)}|^2  \approx 0$  and $|U_{e3(\mu3)}|^2  \approx 0$  is generically reached in the IR, unless the RG flow extends over a too small range of scales \footnote{Quantitatively, this condition reads ${\rm log}_{10}\left(\frac{k_{\rm UV}}{k_{\rm IR}}\right) \ll  y_h^{-2} \frac{\Delta m_{ij}^2}{\Sigma m_{ij}^2}$.}.
For the CKM matrix, $\Sigma m_{ij}^2/\Delta m_{ij}^2 \,\cdot y_h^2\sim \mathcal{O}(1)$ must hold over a sufficiently large range of scales, whereas for the PMNS matrix, the equivalent condition must be violated, i.e., $\Sigma m_{ij}^2/\Delta m_{ij}^2 \,\cdot y_h^2\ll 1$ at all scales.

Thus, we can sharpen our requirements for alternative b): 
First, the tau-Yukawa must not be large over a large range of scales.
Second, the overall mass scale for Dirac neutrinos must not be too large. 
 Only then can the factor $\Sigma m_{ij}^2/\Delta m_{ij}^2 \cdot y_{\tau}$  remain small enough, given the measured difference $\Delta m_{12}^2 \sim  (7.5\pm 0.2) \cdot 10^{-5}{\rm eV^2}$ \cite{deSalas:2020pgw}.

At this point, it may seem that we have collected too many requirements
for such a UV complete theory to exist. 

\section{Asymptotically safe Standard Model with gravity }
Asymptotically safe gravity  \cite{Weinberg:1980gg,Reuter:1996cp}, reviewed in \cite{Eichhorn:2018yfc,Reuter:2019byg,Reichert:2020mja,Eichhorn:2020mte,Bonanno:2020bil,Pawlowski:2020qer,Knorr:2022dsx,Eichhorn:2022jqj,Eichhorn:2022gku,Eichhorn:2023xee,Saueressig:2023irs,Pawlowski:2023gym,Basile:2024oms}, exhibits mechanisms which  -- contingent upon its near-perturbative nature, see \cite{Falls:2013bv,Falls:2014tra,Falls:2017lst,Falls:2018ylp,Eichhorn:2018akn,Eichhorn:2018ydy,Eichhorn:2018nda,Eichhorn:2022gku} and p.~22 in~\cite{Eichhorn:2020sbo}--
achieve all four  requirements
for alternative b), namely 
\begin{enumerate}
\item[i)] a long UV regime beyond the Planck scale \cite{Souma:1999at,Reuter:2001ag,Litim:2003vp,Codello:2008vh,Benedetti:2009rx,Niedermaier:2009zz,Manrique:2011jc,Dona:2013qba,Christiansen:2014raa,Becker:2014qya,Meibohm:2015twa,Gies:2016con,Denz:2016qks,Biemans:2017zca,Wetterich:2019zdo,Burger:2019upn,Fehre:2021eob,Sen:2021ffc,
DAngelo:2023wje,Baldazzi:2023pep,Kluth:2024lar,Falls:2024noj,Saueressig:2025ypi,DAngelo:2025yoy},
\item[ii)] a constant top Yukawa coupling \cite{Eichhorn:2017ylw, Eichhorn:2018whv}, 
\item[iii)] a tau Yukawa coupling that increases from zero 
towards its measured value \cite{ATLAS:2015xst,CMS:2017zyp,ATLAS:2016neq} in the IR \cite{Eichhorn:2022vgp, Kowalska:2022ypk, Pastor-Gutierrez:2022nki}, 
\item[iv)] neutrino Yukawa couplings which can remain tiny
~\cite{Held:2019vmi,Eichhorn:2022vgp, Kowalska:2022ypk,deBrito:2025ges}.
\end{enumerate}
Prior to this work, it was established in \cite{Pendleton:1980as,Alkofer:2020vtb} that the mixing matrices have IR attractive fixed points. Mechanisms to realize the prerequisites (i)-(iv) have been studied in simplified single-generation settings~\cite{Eichhorn:2017ylw, Eichhorn:2018whv,Held:2019vmi,Eichhorn:2022vgp}. Here, we combine these findings and discover a fixed-point cascade that realizes these mechanisms in the full SM. Concretely,
requirement
i) is  achieved, because
quantum-gravity fluctuations  
contribute to all SM  
 RG flows beyond the Planck scale $M_{\rm Planck}$.
We parameterize  quantum-gravity effects by $f_y$ for Yukawa couplings, $f_g$ for gauge couplings and $f_{\lambda}$ for the Higgs quartic coupling, cf.~\cite{Eichhorn:2018whv}. 
Their dependence on the gravitational couplings has been calculated in \cite{Oda:2015sma,Eichhorn:2016esv,Eichhorn:2017eht,Hamada:2017rvn,deBrito:2022vbr}, \cite{Daum:2009dn,Daum:2010bc,Harst:2011zx,Folkerts:2011jz,Christiansen:2017cxa,Eichhorn:2017lry,Christiansen:2017gtg,DeBrito:2019gdd,Eichhorn:2019yzm,Eichhorn:2021qet,deBrito:2022vbr,Pastor-Gutierrez:2022nki} and \cite{Narain:2009fy,Labus:2015ska,Oda:2015sma,Percacci:2015wwa,Hamada:2017rvn,Eichhorn:2017als,Eichhorn:2017ylw,Pawlowski:2018ixd,Wetterich:2019rsn,Eichhorn:2020sbo,Eichhorn:2025ezh}, respectively. 

At the heart of the UV completion is the physics of Yukawa couplings. Their RG evolution in the relevant limit 
is captured by
\begin{align}\label{eq:betaYukawa}
    \beta_{y_i} &=
    -f_y\, y_i
    +\frac{1}{16\pi^2}\Bigg[ 
        3\,\left(
            y_t^2 
            + y_b^2
        \right)y_i
        + \frac{3}{2}\left(
            \delta_{ti}
            + \delta_{bi}
        \right)\,y_i^3
\\&\quad
       - \frac{3}{2}\left(
            |V_{ti}|^2\,y_t^2
            + |V_{ib}|^2\,y_b^2
        \right) y_i   
        - C_\text{U(1)}\,g_Y^2\, y_i
    \Bigg].\notag
\end{align}
For the up-type quarks, $C_{U(1)} = \frac{17}{12}$.
For the down-type quarks, $C_{U(1)} = \frac{5}{12}$.
For the up-type leptons (neutrinos), $C_{U(1)} = \frac{3}{4}$.
For the down-type leptons, $C_{U(1)} = \frac{5}{4}$.
The index $i$ runs over all quark and lepton flavors and we implicitly assume $|V_{ti}|=0$ and $|V_{ib}|=0$, if the respective matrix element does not exist. 

The gravitational contribution $f_y$ is the same for all Yukawa couplings, because gravity is ``flavor-blind". 
 Eq.~\eqref{eq:betaYukawa} has
two fixed points for each Yukawa coupling, namely a free fixed point, $y_{i,\, \rm AF}=0$, and an interacting fixed point, $y_{i\, \rm AS}\neq 0$. If this second fixed point is real, $y_{i,\, \rm AS}$ acts as an upper bound for the Planck-scale value of $y_i$ \cite{Eichhorn:2017ylw, Eichhorn:2018whv}. 

An analogous structure holds for gauge couplings, such that the Abelian gauge coupling is constant at $g_{Y\, \rm AS}= 0.47$ in the asymptotically safe fixed-point regime, if we fix $f_g=9.749 \cdot10^{-3}$ \cite{Eichhorn:2017lry}. The non-Abelian gauge couplings are asymptotically free \cite{Daum:2009dn, Daum:2010bc, Folkerts:2011jz}. 

On the basis of this general structure, we investigate  requirements
ii)-iv), supported by numerical integrations of the full 1-loop beta functions with parameterized gravitational contributions for all 24 SM couplings. 
 
\begin{figure*}[t!]
    \centering
    \includegraphics[width=\linewidth, clip=true, trim=0cm 2.7cm 0cm 4cm]{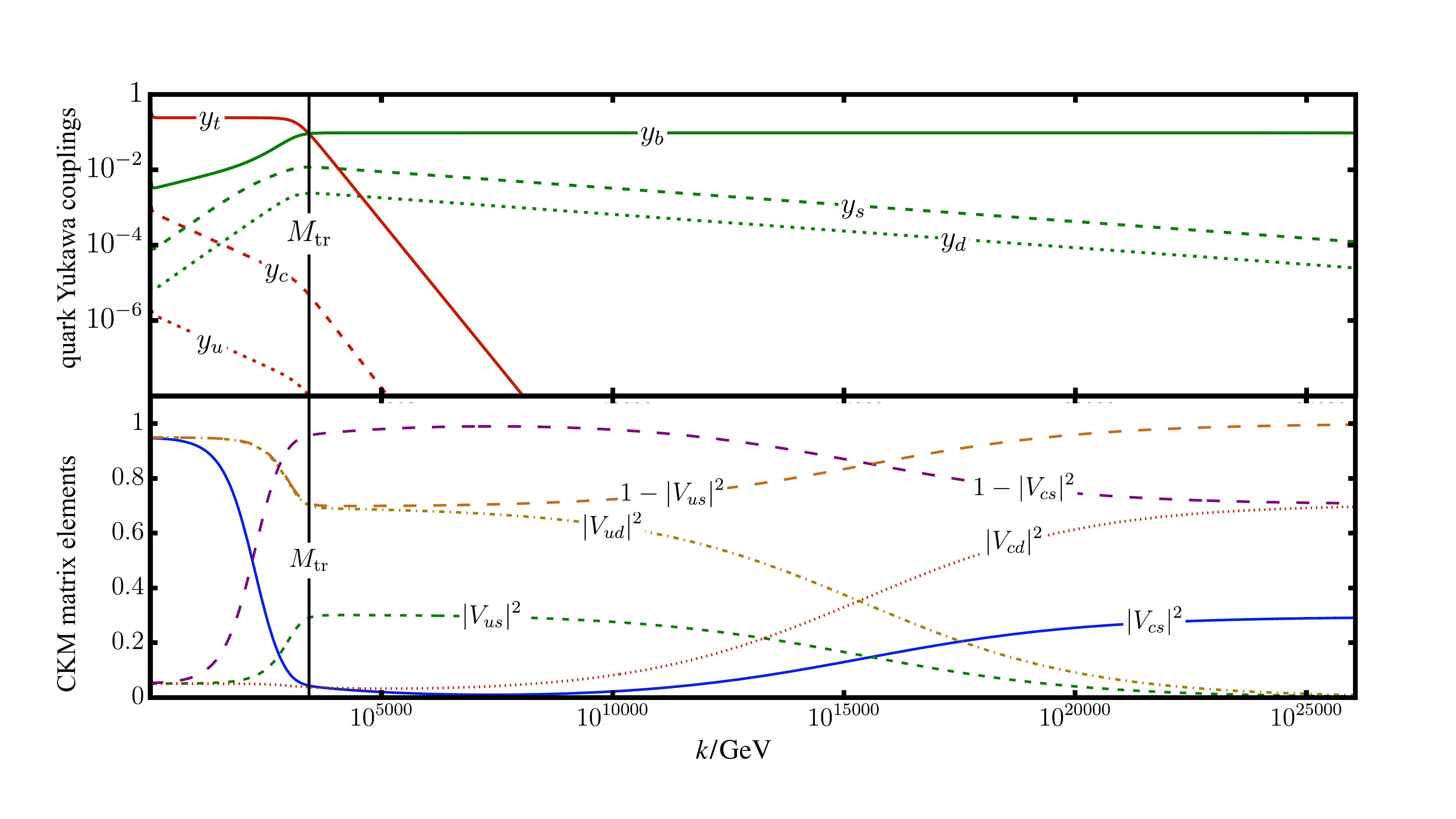}
    \caption{The RG flow for the quark Yukawa couplings (upper panel) and CKM elements (lower panel) transitions from a bottom-dominated regime into a top-dominated regime at $M_{\rm tr} \approx 10^{3400} \rm GeV$. 
    $M_{\rm tr}$
    differs from the leading-order estimate of $1000$ orders of magnitude in scales, because the fixed-point value of the top Yukawa is $y_{t,\,\rm AS}\approx 0.25$, compared to the Planck-scale value of $y_{t}(M_{\rm Planck})\approx 0.38$. The relation $|V_{ud}|^2=1 - |V_{us}|^2$ is already approached in the bottom-dominated regime, while the relation $|V_{cd}|^2=1 - |V_{cs}|^2$ is only approached in the top-dominated regime. This explains why the experimental data show that the first relation holds with higher accuracy than the second one.}
    \label{fig:quarkflows}
\end{figure*}

\subsection{The fixed-point cascade}

\begin{figure*}[t!]
\centering
\includegraphics[width=1\linewidth]{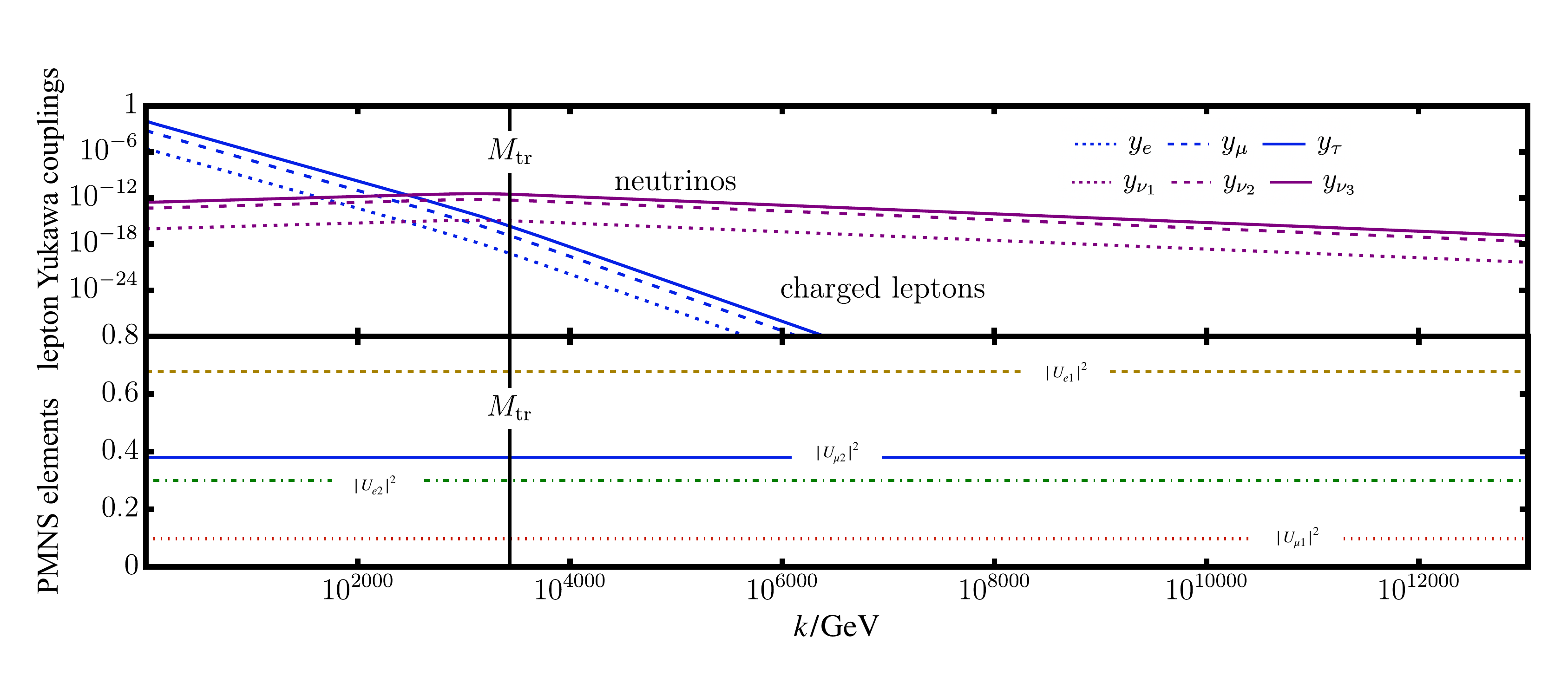}
    \caption{The RG flow for the lepton Yukawa couplings (upper panel) and the PMNS elements (lower panel) differs markedly from the quark sector, see Fig.~\ref{fig:quarkflows}. This results in naturally small neutrino masses as well as a constant PMNS matrix that successfully keeps its non-diagonal, UV configuration intact all the way to the IR.}
    \label{fig:Leptonflows}
\end{figure*}

\begin{figure}[t!]
\centering
\includegraphics[width=0.65\linewidth]{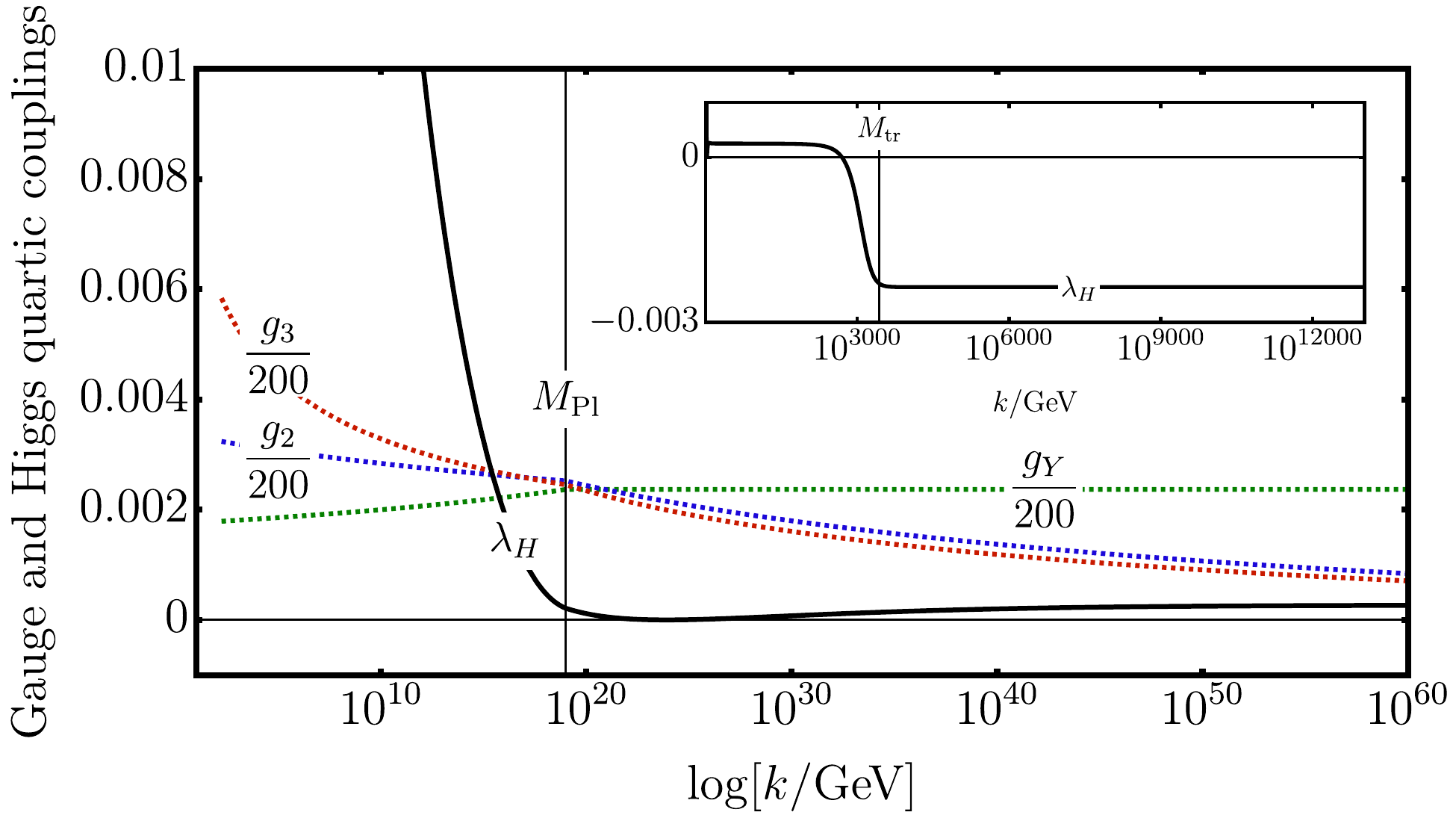}
\caption{We show the RG flow of the gauge-Higgs sector close to the Planck scale. The gauge couplings are multiplied by small numbers for the purposes of this plot. The inset shows a much larger range of scales, over most of which the non-Abelian gauge couplings are completely negligible.}
\label{fig:higgsquarticflow}
\end{figure}

The resulting Figs.~\ref{fig:quarkflows}-\ref{fig:higgsquarticflow} showcase a fixed-point cascade, starting from a true UV completion of the SM. The fixed-point cascade consists of three stages: First, a bottom-dominated fixed-point regime in the very deep UV, second, 
% an intermediate,
top-dominated regime at $M_{\rm Planck}< k< M_{\rm tr}$, with $M_{\rm tr}  \approx 10^{3400}\, \rm GeV$, and third, a SM-regime with the well-established RG flow in the SM without gravity, see, e.g., \cite{Bezrukov:2012sa, Buttazzo:2013uya}. We fix $f_y = -3.27\cdot 10^{-4}$, such that a top pole mass of $M_t \approx 171\, \rm GeV$ (see Appendix \ref{app:Dmatching} for the matching prescriptions) is achieved. The values of $f_y$ and $f_g$ may be understood as a prediction for quantum gravitational fixed-point values \cite{Dona:2013qba,Meibohm:2015twa,Eichhorn:2015bna,Biemans:2017zca,Alkofer:2018fxj,Wetterich:2019zdo,Burger:2019upn,Sen:2021ffc,Pastor-Gutierrez:2022nki}.

The fixed-point cascade owes its existence to  
the interplay of top- and bottom-Yukawa couplings with the Abelian hypercharge. Together, they determine the fixed-point properties for all Yukawa couplings and mixing matrices, cf.~Eq.~\eqref{eq:betaYukawa}. In the top-dominated regime -- required to explain the experimental data on the CKM configuration, cf.~Eq.~\eqref{eq:IRline} -- only some Yukawa couplings are asymptotically free; others, including $y_b$, would vanish in the IR if the top-dominated regime would extend over $k \gg M_{\rm tr}$, contradicting experiment \cite{ATLAS:2018kot,CMS:2018nsn}. Thus, this regime -- while constituting a \emph{UV extension} of the SM  -- cannot be a true UV completion \footnote{Previous studies, when neglecting mixing, suggested a top-dominated regime as a UV completion \cite{Eichhorn:2017ylw, Eichhorn:2018whv}; when including mixing, a top-dominated regime (at nonzero hypercharge coupling) was ruled out as a UV completion because it results in a top pole mass that is too large to be compatible with observations \cite{Alkofer:2020vtb}.}.
Instead, we encounter a bottom-dominated regime, in which all eleven other Yukawa couplings are asymptotically free. It is a \emph{true UV completion of the SM}.

 Hence, requirements i) and ii) hold; the associated huge range of scales challenges expectations regarding the scale of physics beyond the SM. Requirement iii) holds due to a positive critical exponent for the tau-Yukawa coupling, $\theta_{\tau}  = f_y + \frac{3}{4}\frac{-4y_{b/t}^2 +5 g_Y^2}{16\pi^2}\approx 5 \cdot 10^{-3}$ in both regimes. This translates into a steady increase of $y_{\tau}$ throughout the fixed-point cascade, so that $y_{\tau}$ reaches its largest value below the Planck scale. Requirement iv) is fulfilled, because the critical exponents for neutrino Yukawa couplings are tiny throughout the fixed-point cascade, $\theta_{\nu} = f_y + \frac{3}{4}\frac{-4y_{b/t}^2 +g_Y^2}{16\pi^2}$,  yielding
 $\theta_{\nu}\approx  
  6\cdot 10^{-4}
 $ in the bottom-dominated regime. 
Below $M_{\rm tr}$, in the top-dominated regime, they even turn negative, $\theta_{\nu}\approx-
 4\cdot 10^{-4}$, driving the $y_{\nu}'s$ towards smaller values. Hence, asymptotic safety dynamically limits the size of neutrino Yukawa couplings \cite{Held:2019vmi, Kowalska:2022ypk, Eichhorn:2022vgp}. Thus, what appears as an ``unnatural"  gap between neutrino masses and the other SM fermion masses, is  simply
a generic consequence of our UV completion, encoded in different critical exponents and corresponding growth rates of distinct Yukawa couplings towards the IR.

Accordingly, we start the RG flows in the deep UV with fixed-point configurations for CKM elements and PMNS elements~\footnote{For the PMNS matrix elements, this is actually a fixed hypersurface: because all lepton Yukawas approach vanishing values at fixed ratios, the RG flow of the PMNS matrix elements vanishes, see \cite{Alkofer:2020vtb}.}. Both UV fixed lines have 3 relevant directions. In the SM-realization of the RG flow, the CKM elements depart from this fixed line along a relevant direction and are generically attracted towards the IR fixed line.

In contrast, the PMNS matrix elements stay at their UV configuration. While the UV configurations of both CKM and PMNS are far from diagonal, only the PMNS matrix can stay in this configuration, see Fig.~\ref{fig:Leptonflows}.

\subsection{Fixed-point cascade and its predictive power}
For a UV fixed point with $n$ irrelevant directions, i.e., $n$ negative critical exponents, $n$ predictions can be made. For our case, one of them is $g_Y$, because its critical exponent is negative all along the cascade. 
For an IR fixed point with $n$ irrelevant directions, $n$ relations between couplings generically hold with an accuracy that increases, the longer the RG stays close to this fixed point. For our case, these include the top-Yukawa coupling, which is irrelevant at the top-dominated fixed point. Thus, its IR value is predicted with an accuracy related to the length of the regime between $M_{\rm trans}$ and $M_{\rm Planck}$. At the same time, because $y_b$ scales towards zero in the top-dominated regime and is irrelevant at the UV fixed point, its IR value is bounded from above. A large difference between top- and bottom pole mass is thus necessarily achieved.
 
A further prediction of our cascade are the relations \eqref{eq:IRline} between the CKM matrix elements  which we expect to hold approximately, as they do.
An independent cross-check of the cascade is given by measurements of  $|V_{ub}|^2=1-|V_{ud}|^2-|V_{us}|^2=1.39 \cdot 10^{-5} \pm 6 \cdot 10^{-7}$ and $|V_{cb}|^2=1.75 \cdot 10^{-3} \pm 6\cdot 10^{-5}$, see \cite{ParticleDataGroup:2024cfk} and \cite{CLEO:2002eqp,Belle:2018ezy,BaBar:2019vpl,LHCb:2020ist,FermilabLattice:2021cdg,Bordone:2021oof,Bernlochner:2022ucr}. The extent to which these elements depart from zero quantifies the 
 range of scales in the top-dominated regime; because $y_{t,\, AS}\approx 1/4$, $M_{\rm tr} \geq 10^{3000}$ is needed to achieve $|V_{cb}|^2 \approx 10^{-3}$.

\paragraph{A faint fingerprint of the fixed-point cascade --}The fixed-point cascade has additional testable consequences for the CKM elements, because $|V_{ud}|^2=1-|V_{us}|^2$ is an IR attractive relation already in the bottom-dominated regime. 
Accordingly, at the onset of the top-dominated regime, $|V_{ud}|^2+|V_{us}|^2=1$  already holds approximately, whereas $|V_{cd}|^2+|V_{cs}|^2=1$ does not. 
As a consequence, 
we expect that the  measured
values 
satisfy $|V_{ud}|^2+|V_{us}|^2=1$ to  significantly higher
accuracy
than $|V_{cd}|^2+|V_{cs}|^2=1$. 
Experimental data satisfy our expectation since
\begin{eqnarray}
|V_{ud}|^2+|V_{us}|^2 \!&=&\! 0.99999\pm 0.00044, \nonumber\\
|V_{cd}|^2+|V_{cs}|^2 \!&=&\! 0.99825 \pm 0.00044.
\label{eq:bottom_imprint}
\end{eqnarray}
Independent measurements of $|V_{ub}|^2$ and $|V_{cb}|^2$ conform to the same expectation and rule out that $|V_{cb}|^2>|V_{ub}|^2$ at the level of 29$\sigma$. They also result in measurable consequences for  lifetimes and branching fractions of diverse B-mesons \cite{ParticleDataGroup:2024cfk}.

\paragraph{The Higgs sector --}The Higgs quartic coupling is an irrelevant coupling all along the fixed-point cascade~\footnote{The Higgs mass parameter remains a relevant parameter, such that we can achieve a Higgs vev of $246 \, \rm GeV$
in agreement with experimental data.}. Thus, its IR value is a prediction, as proposed in \cite{Shaposhnikov:2009pv}, see also \cite{Eichhorn:2017als,Pawlowski:2018ixd, Wetterich:2019rsn,Eichhorn:2020sbo, Eichhorn:2021tsx,Ohta:2021bkc} and fixes $f_{\lambda} = -5.31 \cdot 10^{-2}$. The resulting value $\lambda_H(M_{\rm Planck}) \approx 0$ corresponds to the stability bound at a top pole mass of $171\, \rm GeV$, corresponding to the measured Higgs with a stable potential below the Planck scale according to \cite{Bezrukov:2012sa,Buttazzo:2013uya,Hiller:2024zjp}, see supp.~material.

The Higgs quartic coupling takes a negative fixed-point value in the deep UV, bottom-dominated, regime, before transitioning to positive values at $M_{\rm trans}$, cf.~Fig.~\ref{fig:higgsquarticflow}. The  slightly negative value may well be lifted above zero by higher-order corrections \cite{Fodor:2007fn,Gies:2014xha,Eichhorn:2015kea,Borchardt:2016xju,Gies:2017zwf} and may not be problematic for the cosmological evolution of our universe, because the very extended
potential barrier results in a low tunneling rate \cite{Coleman:1977py,Callan:1977pt,Sher:1988mj,Casas:1994qy,Espinosa:2007qp,Bezrukov:2012sa,Degrassi:2012ry,Buttazzo:2013uya}.

\section{Conclusion}

We combine the SM with a parameterization of asymptotically safe quantum gravity
with just three free parameters, $f_g$, $f_y$ and $f_{\lambda}$. We can fix these parameters such that we obtain a fixed-point cascade as a UV completion that agrees with
experimental data for SM couplings within 3 $\sigma$ and within the matching prescription detailed in the supplementary material. 
Among these couplings, $g_Y$, $y_t$ and $\lambda_H$ are accurate predictions of the cascade which we use to set $f_g$, $f_y$ and $f_{\lambda}$. In addition, $y_b$ is bounded from above, resulting in a significant gap between top and bottom mass. Both, the fact that $|V_{ud}|^2+|V_{us}|^2=1$ and $|V_{cd}|^2+ |V_{cs}|^2=1$ hold approximately, and that $|V_{ud}|^2+|V_{us}|^2=1$ holds with significantly higher accuracy than $|V_{cd}|^2 + |V_{cs}|^2=1$, are consequences of the cascade. 

Finally, neutrino Yukawa couplings are limited in their size. 
Even at the current upper bound on the effective electron neutrino mass, $m_{\nu_e}^{\rm eff} = \sqrt{\sum_{i =  1,2,3} |U_{e i}|^2 m_i^2} \lesssim 0.45 \, \rm eV$ from the KATRIN experiment \cite{KATRIN:2021uub,KATRIN:2024cdt},  we already obtain a 10\% change of the PMNS elements, see supp.~material. At the significantly lower neutrino mass suggested by cosmological observations \cite{DESI:2024mwx,DESI:2025ejh}, there are only sub-percent changes.
We therefore work with an effective electron neutrino mass of $m_{\nu_{e}}^{\rm eff}\approx 0.009\,\rm eV$ throughout this paper, at which the PMNS configuration is constant, 
cf.~Fig.~\ref{fig:Leptonflows}.
Throughout our fixed-point cascade, the requirement of low enough Yukawa couplings is met. The observed large gap in the spectrum of fermion masses in the SM is both a prerequisite for significant neutrino mixing as well as a natural consequence within our fixed-point cascade.

In fact, lepton mixing would already be driven towards zero for a neutrino mass scale higher than $\sim \mathcal{O}(1-10)\, \rm eV$ for the RG flow between $10^{19}\, \rm GeV$ and the electroweak scale for both normal and inverted ordering. Irrespective of the UV completion, this rules out larger neutrino masses; this
is subject to a forthcoming publication \cite{ZoisAaronAstridtoappear}. We also expect that seesaw mechanisms may be in tension with our mechanism, see \cite{ZoisAaronAstridtoappear2}.

At a more abstract level, we learn that RG flows over huge ranges of scales -- traditionally met with skepticism -- may after all
be necessary to understand the origin of structures in the SM. Long ranges of scales are an automatic consequence of asymptotically safe scaling regimes, because a true scaling regime implies that an arbitrarily large range of scales is described by the fixed-point properties. In most studies to date, this was note made explicit, because the critical exponents are often larger than those in our system. Here, the tiny values of the critical exponents in our system directly translate into large ranges of scales which are necessary in order to track the resulting changes in the values of couplings.
In fact,
these huge ranges of scales are %in fact 
imprinted in the experimental data, through the critical exponents of CKM matrix elements, together with the two relations which are satisfied to high, but -- crucially -- different degrees of accuracy.

%{This holds under the assumption that beyond-SM physics, required to explain dark matter and the matter-antimatter asymmetry, is only very weakly coupled to the SM, so that the resulting RG ƒflows are not too different from those we present here.}

As an outlook, we add that the present setting needs to be extended by a phenomenologically viable dark-matter scenario as well as a mechanism to explain the matter-antimatter asymmetry. It is already known that dark-matter models can be severely constrained or even ruled out when coupled to asymptotically safe gravity \cite{Eichhorn:2017als, Reichert:2019car,Eichhorn:2020sbo,Kowalska:2020zve,Hamada:2020vnf,Eichhorn:2021tsx,deBrito:2021akp,Boos:2022pyq,deBrito:2023ydd,Assant:2025gto}. Studying beyond SM settings can be done along the lines we have pursued here, by adding the quantum gravitational effects to any new sectors in the same parameterized form, as in \cite{Reichert:2019car,Kowalska:2020zve,Hamada:2020vnf,deBrito:2023ydd}. 
Very weakly coupled dark sectors as well as primordial black holes might constitute viable candidates, because the corresponding new physics degrees of freedom are coupled weakly enough to the SM that they should not significantly impact the mechanisms that we have discovered. For the case of the matter-antimatter asymmetry, the situation may be more subtle, because additional degrees of freedom may not be sufficiently weakly coupled. Along a different direction, quantum gravity may itself constitute a source of parity and/or CP violation \cite{Borissova:2025frj} and thereby offer new mechanisms to contribute to matter-antimatter asymmetry.

\section*{Acknowledgements} 

We gratefully acknowledge discussions and comments on the manuscript by  Gustavo de Brito, Holger Gies, Jan Pawlowski, Manuel Reichert and Marc Schiffer.
A.~E.~would like to thank the Laboratoire de Physique de l'École normale superieure for hospitality. A.~H.~would like to thank the Institute for Theoretical Physics at Heidelberg University for hospitality. This work is funded by the Deutsche Forschungsgemeinschaft (DFG, German Research Foundation) under Germany’s Excellence Strategy EXC 2181/1 - 390900948 (the Heidelberg STRUCTURES Excellence Cluster).
A.~E.~acknowledges the European Research Council's (ERC) support under the European Union’s Horizon 2020 research and innovation program Grant agreement No.~101170215 (ProbeQG). 

\begin{appendix}
\numberwithin{equation}{section}

\section{Beta functions}

We provide the 1-loop beta functions of the Standard Model couplings extended with three Dirac right-handed neutrinos. We denote as $\{g_3, g_2, g_Y\}$ the ${\rm SU}(3)$, ${\rm SU}(2)_L$ and ${\rm U}(1)_Y$ gauge couplings, $\{y_t, y_b, y_c,y_s, y_u,y_d\}$ the quark Yukawa couplings, $\{y_{\nu_3}, y_{\tau}, y_{\nu_2,}y_{\mu}, y_{\nu_1},y_e\}$ the lepton Yukawa couplings , and $\lambda_H$ the Higgs quartic coupling. We also provide the 1-loop beta functions for the CKM and PMNS matrix elements. The 1-loop beta functions for the gauge couplings are:
\begin{eqnarray}
   &\beta_{g_y} = -f_g g_y + \frac{41}{6}\frac{g_y^3}{\left(16 \pi ^2\right)} \;\;,\\
   &\beta_{g_2} = -f_g g_2 - \frac{19}{6}\frac{g_2^3}{16 \pi ^2} \;\;,\\
   &\beta_{g_3} = -f_g g_3 - \frac{7}{6}\frac{g_3^3}{16 \pi ^2}  \;. \;\;
\end{eqnarray}

The 1-loop beta functions of the up-type quark Yukawa couplings ($i = u,c,t$) are:

\begin{eqnarray}
    \beta_{y_i} &=& +\frac{y_i}{16\pi^2} \bigg[ \frac{3}{2}y_i^2 + \left(3 \sum_{j}y_j^2 + 3 \sum_{\kappa} y_{\kappa}^2 + \sum_{l}y_l^2 + \sum_{k} y_{\nu_k}^2 \right) \nonumber \\ 
    && -\left(\frac{17}{12} g_Y^2 +\frac{9}{4}g_2^2 +8 g_s^2\right) - \frac{3}{2} \sum_{\kappa} y_\kappa^2 |V_{i\kappa}|^2\bigg]\label{eq:betaYukawa1}\,\,,
    \end{eqnarray}

while for the down-type quarks ($\kappa = d,s,b$) one has: 
\begin{eqnarray}
    \beta_{y_\kappa} &=& +\frac{y_\kappa}{16\pi^2} \bigg[ \frac{3}{2}y_\kappa^2 + \left(3 \sum_{j}y_j^2 + 3 \sum_{\kappa} y_{\kappa}^2 + \sum_{l}y_l^2 + \sum_{k} y_{\nu_k}^2 \right)\nonumber  \\ 
    && -\left(\frac{5}{12} g_Y^2 +\frac{9}{4}g_2^2 +8 g_s^2\right) - \frac{3}{2} \sum_{i} y_i^2 |V_{i\kappa}|^2\bigg]\label{eq:betaYukawa2}\,\,,
    \end{eqnarray}

    Analogously the beta functions for the Yukawa couplings of the up-type lepton , i.e., the neutrinos ($\nu_i = \nu_1,\nu_2,\nu_3 $), are:
    \begin{eqnarray} 
    \beta_{y_{\nu_i}} &=&\frac{y_{\nu_l}}{16\pi^2} \bigg[\frac{3}{2}y_{\nu_l}^2 +3 \sum_{j}y_j^2 + 3 \sum_{\kappa} y_{\kappa}^2 + \sum_{l}y_l^2 + \sum_{k} y_{\nu_k}^2 \nonumber \\
    && -\left(\frac{3}{4} g_Y^2 +\frac{9}{4}g_2^2 
    \right) -\frac{3}{2} \sum_{l} y_{l}^2 |U_{l i}|^2\bigg]\,\,,
\end{eqnarray}

and for the down-type leptons ($l = e,\mu,\tau$):

\begin{eqnarray}
    \beta_{y_l} &=&\frac{y_l}{16\pi^2} \bigg[\frac{3}{2}y_l^2 +3 \sum_{j}y_j^2 + 3 \sum_{\kappa} y_{\kappa}^2 + \sum_{l}y_l^2 + \sum_{k} y_{\nu_k}^2 \nonumber  \\ 
    && -\left(\frac{5}{4} g_Y^2 +\frac{9}{4}g_2^2
    \right) -\frac{3}{2} \sum_{k} y_{\nu_k}^2 |{U}_{l k}|^2\bigg]. 
\end{eqnarray}

The 1-loop beta function of the Higgs quartic coupling is given as:
\begin{eqnarray}
    \beta_{\lambda_H} &=&\frac{1}{16 \pi^2} \Bigg[24 \lambda_H^2 - 3 \lambda_H \left(3 g_2^2+g_y(t)^2\right) +\frac{6}{16}\left(2 g_2^2 g_y(t)^2+3 g_2^4+g_y(t)^4\right)  \nonumber\\
    &{}&+ 4 \lambda_H \left(3 \sum_{\rm quarks} y_i^2 + \sum_{\rm leptons} y_l^2+ \sum_{\rm neutrinos} y_{\nu_l}^2 \right) \nonumber \\ 
    &&- 2 \left(3 \sum_{\rm quarks} y_i^4 + \sum_{\rm leptons} y_l^4 + \sum_{\rm neutrinos} y_{\nu_l}^4\right)\Bigg].
\end{eqnarray}

The gravitational contributions are parameterized by a single parameter $f_g$ for all three gauge couplings. This follows,  because gravity couples to energy-momentum, and is ``blind" to internal symmetries. Further, explicit calculations show that the gravitational contribution is linear in the gauge coupling. Similarly, there is one linear contribution with coefficient $f_y$ for all the Yukawa couplings and a linear contribution $f_{\lambda}$ for the quartic coupling. 

Finally, below the Planck scale, the gravitational contribution drops to zero quadratically in $k$ and is thus quickly negligible compared to the other terms in the beta function, which only change logarithmically. We approximate this by setting the gravitational contribution to zero at the Planck scale. It has been checked explicitly that this is a good approximation, see \cite{Eichhorn:2017ylw,Kotlarski:2023mmr}.  These contributions can also be calculated from first-principles with functional RG techniques \cite{Wetterich:1992yh,Morris:1993qb}, reviewed in \cite{Dupuis:2020fhh}. In fact, in our phenomenological parameterization, they may be understood as placeholders for an effective resummation of quantum-gravitational effects.

In contrast, gravitational contributions do not directly enter the beta functions of the CKM and PMNS elements. This is because the reparametrization of the Standard Model fermions from the flavor basis to the mass basis is indistinguishable for gravity which couples to energy-momentum. The mixing information for the quarks is encoded in the CKM matrix which can be parametrized with four independent elements. The PMNS matrix, encoding the mixing information for the leptons, also contains four independent parameters assuming the existence of only 3 generations of Dirac neutrinos. In our model, we assume the unitarity of both the CKM and the PMNS matrices which upon using the following parametrization (for the squared matrices):
\begin{eqnarray}
|V|^2=\left(
\begin{matrix}
X & Y & 1 - X - Y\\
Z & W & 1 - W - Z\\
1 - X - Z & 1- Y - W& W + X + Y + Z -1
\end{matrix}\right).
\end{eqnarray}

The flow equation of the above matrix elements are given as follows:
\newline
\resizebox{1\linewidth}{!}{
\centering
$
\begin{aligned}
\beta_X&=-\frac{3}{(4\pi)^{2}}\left[\frac{y^{2}_{u (e)}+y^{2}_{c(\mu)}}{y^{2}_{u (e)}-y^{2}_{c(\mu)}}\left\{(y^{2}_{d (\nu_1)}-y^{2}_{b(\nu_3)})XZ +\frac{y^{2}_{b(\nu_3)}-y^{2}_{s (\nu_2)}}{2}(W(1-X)+X-(1-Y)(1-Z))\right\}\right.
\label{eq:betaX} 
\\
&\left.\phantom{\frac{1}{2}} +\frac{y^{2}_{u (e)}+y^{2}_{t(\tau)}}{y^{2}_{u (e)}-y^{2}_{t(\tau)}}\left\{(y^{2}_{d (\nu_1)}-y^{2}_{b(\nu_3)})X(1-X-Z)+\frac{(y^{2}_{b(\nu_3)}-y^{2}_{s (\nu_2)})}{2}((1-Y)(1-Z)-X(1-2Y)-W(1-X)) \right\} \right. 
\nonumber \\
&\left.\phantom{\frac{1}{2}} +\frac{y^{2}_{d (\nu_1)}+y^{2}_{s (\nu_2)}}{y^{2}_{d (\nu_1)}-y^{2}_{s (\nu_2)}}\left\{ (y^{2}_{u (e)}-y^{2}_{t(\tau)})XY+\frac{y^{2}_{t(\tau)}-y^{2}_{c(\mu)}}{2}(W(1-X)+X-(1-Y)(1-Z)) \right\} \right. 
\nonumber\\
 &\left.\phantom{\frac{1}{2}} +\frac{y^{2}_{d (\nu_1)}+y^{2}_{b(\nu_3)}}{y^{2}_{d (\nu_1)}-y^{2}_{b(\nu_3)}}\left\{(y^{2}_{u (e)}-y^{2}_{t(\tau)})X(1-X-Y)+\frac{y^{2}_{t(\tau)}-y^{2}_{c(\mu)}}{2}((1-Y)(1-Z)-X(1-2Z)-W(1-X))\right\} \right],
\nonumber
\end{aligned}
$
}
\newline
\resizebox{\linewidth}{!}{
\
$
 \begin{aligned}
   \scriptsize\beta_Y&=-\frac{3}{(4\pi)^{2}}\left[\frac{y^{2}_{u(e)}+y^{2}_{c(\mu)}}{y^{2}_{u (e)}-y^{2}_{c(\mu)}}\left\{\frac{y^{2}_{b(\nu_3)}-y^{2}_{d(\nu_1)}}{2}\left(W(1-X)+X-(1-Y)(1-Z)\right)+\left(y^{2}_{s (\nu_2)}-y^{2}_{b(\nu_3)}\right)YW\right\}\right. \nonumber \\
    &\left.\phantom{\frac{1}{2}} +\frac{y^{2}_{u (e)}+y^{2}_{t(\tau)}}{y^{2}_{u (e)}-y^{2}_{t(\tau)}}\left\{ \frac{\left(y^{2}_{b(\nu_3)}-y^{2}_{d (\nu_1)}\right)}{2}\left((1-Y)(1-Z)-W(1-X)-X(1-2Y)\right)+\left(y^{2}_{s (\nu_2)}-y^{2}_{b(\nu_3)}\right)Y(1-Y-W) \right\} \right. \nonumber\\
    &\left.\phantom{\frac{1}{2}} +\frac{y^{2}_{s (\nu_2)}+y^{2}_{d (\nu_1)}}{y^{2}_{s (\nu_2)}-y^{2}_{d (\nu_1)}}\left\{\left(y^{2}_{u (e)}-y^{2}_{t(\tau)}\right)XY+\frac{y^{2}_{t(\tau)}-y^{2}_{c(\mu)}}{2}\left(W(1-X)+X-(1-Y)(1-Z)\right)\right\} \right. \nonumber \\
    & \left.\phantom{\frac{1}{2}} +\frac{y^{2}_{s (\nu_2)}+y^{2}_{b(\nu_3)}}{y^{2}_{s (\nu_2)}-y^{2}_{b(\nu_3)}}\left\{(y^{2}_{u (e)}-y^{2}_{t(\tau)})Y(1-X-Y)
    +\frac{(y^{2}_{c(\mu)}-y^{2}_{t(\tau)})}{2}(W(1-X-2Y)+X-(1-Z)(1-Y)) \right\}\right],
     \label{eq:betaY}
\end{aligned}
$
}
\newline
\resizebox{\linewidth}{!}{
\
$
 \begin{aligned}
  \beta_Z&=-\frac{3}{(4\pi)^{2}}\left[\frac{y^{2}_{c(\mu)}+y^{2}_{u (e)}}{y^{2}_{c(\mu)}-y^{2}_{u (e)}}\left\{(y^{2}_{d (\nu_1)}-y^{2}_{b(\nu_3)})XZ+\frac{(y^{2}_{b(\nu_3)}-y^{2}_{s (\nu_2)})}{2}(W(1-X)+X-(1-Z)(1-Y))\right\}  \right. \nonumber\\
      & \left.\phantom{\frac{1}{2}}+\frac{y^{2}_{c(\mu)}+y^{2}_{t(\tau)}}{y^{2}_{c(\mu)}-y^{2}_{t(\tau)}}\left\{(y^{2}_{d (\nu_1)}-y^{2}_{b(\nu_3)})Z(1-X-Z)
         +\frac{(y^{2}_{s (\nu_2)}-y^{2}_{b(\nu_3)})}{2}(W(1-X-2Z)+X-(1-Y)(1-Z))\right\} \right. \nonumber\\
         &\left.\phantom{\frac{1}{2}} +\frac{y^{2}_{d (\nu_1)}+y^{2}_{s (\nu_2)}}{y^{2}_{d (\nu_1)}-y^{2}_{s (\nu_2)}}\left\{\frac{(y^{2}_{u (e)}-y^{2}_{t(\tau)})}{2}((1-Y)(1-Z)-X-W(1-X))+(y^{2}_{c(\mu)}-y^{2}_{t(\tau)})ZW\right\} \right. \nonumber \\
     & \left.\phantom{\frac{1}{2}}+\frac{y^{2}_{d (\nu_1)}+y^{2}_{b(\nu_3)}}{y^{2}_{d (\nu_1)}-y^{2}_{b(\nu_3)}}\left\{\frac{(y^{2}_{t(\tau)}-y^{2}_{u (e)})}{2}((1-Z)(1-Y)-W(1-X)-X(1-2Z))+(y^{2}_{c(\mu)}-y^{2}_{t(\tau)})Z(1-Z-W) \right\}\right],
      \label{eq:betaZ}
\end{aligned}
$
}
\newline
\resizebox{\linewidth}{!}{
\centering
$
 \begin{aligned}
 \beta_W&= -\frac{3}{(4\pi)^{2}}\left[\frac{y^{2}_{c(\mu)}+y^{2}_{u (e)}}{y^{2}_{c(\mu)}-y^{2}_{u (e)}}\left\{(y^{2}_{s (\nu_2)}-y^{2}_{b(\nu_3)})WY+\frac{(y^{2}_{b(\nu_3)}-y^{2}_{d (\nu_1)})}{2}((1-X)W+X-(1-Y)(1-Z))\right\}  \right. \nonumber \\
     &  \left.\phantom{\frac{1}{2}} +\frac{y^{2}_{c(\mu)}+y^{2}_{t(\tau)}}{y^{2}_{c(\mu)}-y^{2}_{t(\tau)}}\left\{(y^{2}_{s (\nu_2)}-y^{2}_{b(\nu_3)})W(1-Y-W)
         +\frac{(y^{2}_{b(\nu_3)}-y^{2}_{d (\nu_1)})}{2}((1-Y)(1-Z)-X-W(1-X-2Z))\right\} \right. \nonumber\\
         &\left.\phantom{\frac{1}{2}} +\frac{y^{2}_{s (\nu_2)}+y^{2}_{d (\nu_1)}}{y^{2}_{s (\nu_2)}-y^{2}_{d (\nu_1)}}\left\{(y^{2}_{c(\mu)}-y^{2}_{t(\tau)})WZ+\frac{(y^{2}_{t(\tau)}- y^{2}_{u (e)})}{2}((1-X)W+X-(1-Y)(1-Z))\right\} \right. \nonumber \\
     & \left.\phantom{\frac{1}{2}}+\frac{y^{2}_{s (\nu_2)}+y^{2}_{b(\nu_3)}}{y^{2}_{s (\nu_2)}-y^{2}_{b(\nu_3)}}\left\{(y^{2}_{c(\mu)}-y^{2}_{t(\tau)})W(1-Z-W)+\frac{(y^{2}_{t(\tau)}-y^{2}_{u (e)})}{2}((1-Y)(1-Z)-X-W(1-X-2Y)) \right\}\right].
      \label{eq:betaW}
\end{aligned}
$
}
\newline 

Beta functions capture the scale dependence of couplings and there are different beta functions for different notions of running. In our context, the distinction between running as a function of the physical momentum and running as a function of an auxiliary regularization scale is most important, and has in the past led to confusion in the interplay of quantum gravity and matter. These two notions of running are in general not the same, see, e.g., \cite{Donoghue:2019clr,Bonanno:2020bil}, see also \cite{Buccio:2024hys} for an explicit demonstration of this fact. If all couplings in the setup in question are dimensionless, there is typically only logarithmic running and in this case, different notions of running track each other and are interchangeable. Because gravity has a dimensionful coupling, the Newton coupling, this is no longer the case, when gravitational fluctuations are involved, and has led to a lengthy debate in the literature, see, e.g., \cite{Robinson:2005fj,Pietrykowski:2006xy,Ebert:2007gf,Toms:2007sk,Toms:2008dq,Tang:2008ah,Rodigast:2009zj,Toms:2010vy,Anber:2010uj,Ellis:2010rw,Toms:2011zza,Eichhorn:2017ylw, Frob:2017lnt,Ferreiro:2018oxx,Eichhorn:2018whv} and references therein. By now, this difference is well-understood. Both notions are well-studied in asymptotic safety, see, e.g., \cite{Knorr:2022lzn,Knorr:2022dsx,Pastor-Gutierrez:2024sbt} for studies of the running with physical momentum in asymptotic safety.

Our focus is on the running as a function of a cutoff/regularization scale, $k$. This notion is relevant to understand which parameters of a theory are predicted and which are free parameters. Therefore, considering $k$-running at highly transplanckian scales is physically meaningful, because it determines whether or not a given coupling is a prediction of an asymptotically safe setting, see also \cite{Saueressig:2024ojx}.
We note in passing that a careful treatment also allows to extract this notion of running from perturbative calculations, and that there is a non-zero gravitational contribution to this type of running \cite{deBrito:2022vbr}.

%%%%%%%%
\section{Derivation of Eq.~9 for the mixing matrix elements}\label{app:Bderivation}
%%%%%%%%
From Eq.~\eqref{eq:betaX}-\eqref{eq:betaW}, we obtain the beta functions  $\beta_{|V_{13}|^2}$ and $\beta_{|V_{23}|^2}$ through
\begin{eqnarray}
\beta_{|V_{13}|^2} &=& \partial_t \left(1-X-Y \right)\nonumber\\
&=&\left(-\beta_X - \beta_Y\right)|_{Y\rightarrow 1-X-|V_{13}|^2, \, Z\rightarrow 1- W- |V_{23}|^2},\\
\beta_{|V_{23}|^2} &=& \partial_t \left(1-W-Z \right)\nonumber\\
&=&\left(-\beta_W - \beta_Z\right)|_{Y\rightarrow 1-X-|V_{13}|^2, \, Z\rightarrow 1- W- |V_{23}|^2}.
\end{eqnarray}

To take the heavy-top limit, it suffices to send all the other quark Yukawa couplings to zero. However, because the mixing metric is ill-defined in the degenerate limit, i.e., $y_i^2 -y_j^2 \to 0$, the limits of vanishing Yukawas must be taken in the phenomenological ordering. That is: i) $y_u \to0$, ii) $y_d \to0$, iii) $y_s\to0$, iv) $y_c\to0$, and lastly 5) $y_b\to0$. 

This results in

\begin{eqnarray}
\beta_{|V_{13}|^2}\!\! &=& \!\!\frac{3\, y_t^2}{16\pi^2} |V_{13}|^2  +\mathcal{O}(y_i^2,|V_{13}|^2,|V_{23}|^2) \label{eq:B3} \;,\;\\
\beta_{|V_{23}|^2} \!\!&=&\!\!\frac{3\, y_t^2}{16\pi^2} |V_{23}|^2 +\mathcal{O}(y_i^2,|V_{13}|^2,|V_{23}|^2) \label{eq:B4}.
\end{eqnarray}

From these expressions, we infer that $|V_{13}|=0=|V_{23}|$ corresponds to an approximate fixed line and we can also read off the critical exponents
\begin{eqnarray}
\theta_{|V_{13}|^2} =-\frac{\partial \beta_{|V_{13}|^2}}{\partial |V_{13}|^2}
\Big|_{|V_{13}|^2=0=|V_{23}|^2}= -\frac{3y_t^2}{16\pi^2} = \theta_{|V_{23}|^2}.
\end{eqnarray}

Let us instead assume that the difference between $y_s$ and $y_b$ is small, $\Delta_{bs}= y_b^2-y_s^2$ . We therefore consider
\begin{equation}
\beta_{|V_{13(23)}|^2} = \frac{1}{\Delta m^2_{bs}}\underset{\Delta m^2_{bs} \rightarrow 0}{\rm lim} \Delta m^2_{bs}\cdot \beta_{|V_{13(23)}|^2}+ \mathcal{O}(\Delta_{sb}).
\end{equation}
Then, to leading order in $|V_{13(23)}|^2$ (where we impose the fixed-line relation in the form $W=1-Z (X = 1 - Y)$, which hold to leading order close to the line), we obtain
\begin{eqnarray}
\beta_{|V_{13}|^2} &=& \frac{3}{32\pi^2} |V_{13}|^2\, y_t^2\, \frac{\Sigma m^2_{bs}}{\Delta m^2_{bs}} \left(1+Y \right) +\dots,\\
\beta_{|V_{23}|^2} &=&\frac{3}{32\pi^2} |V_{23}|^2\, y_t^2\,\frac{\Sigma m^2_{bs}}{\Delta m^2_{bs}} \left(1-Y \right)+\dots,
\end{eqnarray}
where we have also introduced the sum $\Sigma_{sb} = y_b^2+y_s^2$. Note that, singling out the specific ratio $\frac{\Sigma m^2_{bs}}{\Delta m^2_{bs}}$ leads to a relevant factor of $\frac{1}{2}$ compared to Eqs.~\eqref{eq:B3}-\eqref{eq:B4}, where all ratios are reduced to unity.  Along the fixed line, the critical exponent depends on the value of $Y$, but close to the endpoint, at which $Y=0$, we obtain the expression in Eq.~9.

\section{Overview of couplings}

While the trajectories that we show in the main text can be hard to find from the IR, they are easily reproducible starting from a set of UV values. To provide an overview over the fixed-point cascade and its predictions, we list the following pieces of information for each coupling below: i) the experimental value of the coupling in the IR, ii) the IR value of the coupling that results from our fixed-point cascade, iii) the associated (approximate) critical exponent in the top-dominated regime
and, iv) the critical exponent at the UV fixed point.

The UV values may be used as initial conditions that enable the reader to reproduce our numerical results, in particular Fig.~1-3.

For CKM and PMNS matrix elements, we use the global fits in \cite{ParticleDataGroup:2024cfk}, which account for unitarity of both mixing matrices.

\begin{table*}
\centering
\begingroup
\setlength{\tabcolsep}{1.4pt} 
\renewcommand{\arraystretch}{1.3} % Default value: 1
\begin{tabular}{|c|c|c|c|c|c|}
coupling & Exp. value & IR (RG)& UV & $\theta_t$   
&$\theta_{UV}$ \\ \hline\hline
$g_Y$ &  $(3.5855\pm 0.0007)\cdot 10^{-1}$ & $0.358518$ & $0.47465$  &$-0.019$ 
& $-0.019$ \\ \hline
$g_2$ &  $(6.4765\pm0.0028) \cdot 10^{-1}$ & fixed in IR & $\sim 0 $& $0.0097$  
& $0.0097$ \\ \hline
$g_3$ &  $1.1618 \pm 0.0045$ & fixed in IR & $\sim 0 $  &  $0.0097$  
& $0.0097$  \\ \hline \hline
$y_t$ &$(9.2437 \pm 0.043)\cdot 10^{-1}$ & $9.212\cdot 10^{-1}  $ &  $4.58839 \cdot 10^{-36}$  & $-0.0034$ 
& $0.0015$\\ \hline
$y_b$ & $(1.55334 \pm 0.14)\cdot 10^{-2}$ & $1.57\cdot10^{-2}  $& $9.68782\cdot10^{-2}$ & $-0.0030$  
& $-0.00053$ \\ \hline
$y_c$ & $(3.4150 \pm 0.0098)\cdot 10^{-3}$ &$3.4082 \cdot 10^{-3}$& $ 2.197\cdot10^{-41}$  & $0.00056$ 
& $0.0015$\\ \hline
$y_s$ & $(2.93 \pm 0.25)\cdot 10^{-4}$ & $2.95\cdot 10^{-4}$ & $1.217\cdot10^{-4}$  &$-0.00086$  
& $0.000089$\\ \hline
$y_u$ & $(6.8 \pm 1.1)\cdot 10^{-6}$ & $6.7595\cdot 10^{-6}$ & $5.864\cdot10^{-44}$  & $0.00056$ 
& $0.0016$ \\ \hline
$y_d$ & $(1.5 \pm 0.1)\cdot10^{-5}$ & $1.4557\cdot 10^{-5}$ & $2.393\cdot10^{-5}$  & $-0.00086$ 
& $0.000089$ \\ \hline \hline
$y_e$ & $(2.7930 \pm0.0016)\cdot 10^{-6}$ & $2.79483\cdot 10^{-6}$ & $16.836\cdot10^{-130}$  & $0.0039$  
& $0.0048$ \\ \hline
$y_{\mu}$ & $(5.8838\pm 0.0011)\cdot 10^{-4}$& $5.88731\cdot 10^{-4}$ & $35.465\cdot10^{-128}$  &$0.0039$ 
& $0.0048$\\ \hline
$y_{\tau}$ & $(9.9944 \pm 0.0008)\cdot10^{-3}$ &$9.9955\cdot10^{-3}$& $6.022\cdot10^{-126}$  &$0.0039$  
& $0.0048$ \\ \hline
$y_{\nu_1}$ &  $\approx 10^{-16}$ (choice)& $1.00074\cdot10^{-16}  $ & $19.075\cdot 10^{-29}$ & $-0.00039$  
& $0.00056$ \\ \hline
$y_{\nu_2}$ &  $(4.97 \pm 0.06)\cdot10^{-14}$ & $4.94316\cdot10^{-14}$& $9.422\cdot10^{-26}$  & $-0.00039$  
& $0.00056$ \\ \hline
$y_{\nu_3}$ &  $(2.89 \pm 0.01 )\cdot10^{-13}$ & $2.87167\cdot 10^{-13}$ & $54.735\cdot10^{-26}$  & $-0.00039$ 
& $0.00056$ \\ \hline \hline
$|V_{cs}|^2$ & $(9.4768 \pm 0.0031)\cdot 10^{-1}$ & $9.47630 \cdot 10^{-1}$ & $2.9101178\cdot10^{-1}$& $ -1.13\cdot 10^{-3}$  
&  $1.78\cdot 10^{-4}\,\,$ \\ \hline
$|V_{ud}|^2$ & $(9.4936\pm 0.0031)\cdot 10^{-1}$ & $9.49280\cdot10^{-1}$ & $8.63112\cdot10^{-3}$& $ -1.13\cdot 10^{-3} \,\,$  
& $1.78\cdot 10^{-4}\,\,$ \\ \hline
 $|V_{us}|^2$ & $(5.063 \pm 0.031)\cdot10^{-2}$ & $5.0711\cdot10^{-2}$ & $3.814289\cdot10^{-3}$& $ -1.13\cdot 10^{-3}$  
 &  $1.78\cdot 10^{-4} \,\,$\\ \hline
$|V_{cd}|^2$ & $(5.057 \pm 0.031)\cdot 10^{-2}$ & $5.06495\cdot10^{-2} $ & $6.9654997\cdot10^{-1}$&  $\sim 0 \,\,$ 
&  $\sim 0$ \\ \hline\hline
$|U_{\mu2}|^2$ & $(3.80\pm0.51) \cdot 10^{-1}$ & $3.799\cdot10^{-1}$ & $3.7986\cdot 10^{-1}$& $\sim 0$ 
& $\sim 0$\\ \hline
$|U_{e1}|^2$ & $(6.78 \pm 0.12 )\cdot 10^{-1}$  & $6.7803\cdot10^{-1}$ & $6.7799\cdot 10^{-1}$& $\sim 0$  
& $\sim 0$ \\ \hline
$|U_{e2}|^2$ & $(3.00\pm 0.12)\cdot 10^{-1}$ & $3.0037\cdot10^{-1}$ & $3.0041\cdot10^{-1}$&$\sim 0$ 
& $\sim 0$ \\ \hline
$|U_{\mu1}|^2$ & $(9.7 \pm 3.8)\cdot10^{-2} $& $9.7615\cdot 10^{-2}$ & $9.761\cdot 10^{-2}$&$\sim 0$  
& $\sim 0$\\ \hline\hline
 & & $0.0002$ &  &  & \\
$\lambda_H$ &$0.12607 \pm 0.0003$ & (at $k=M_{\rm Planck}$ & $-2.37324\cdot 10^{-3}$ & $-0.053$ & $-0.049$\\ 
& & $\lambda_H \approx 0$) & & & \\ \hline \hline
\end{tabular}
\caption{The quoted IR values correspond to top pole-mass scale $M_t = 171\, \rm GeV$. 
The experimental values for all gauge couplings and Yukawa couplings have been taken from \cite{Huang:2020hdv} corresponding to $M_t = 173.1\,\rm GeV$. We assume that the running for all but the top Yukawa is negligible between $173.1\, \rm GeV$ and $171\,\rm GeV$. The top Yukawa has been calculated by applying the 2-loop matching prescription described in \cite{Buttazzo:2013uya}. The experimental values of the mixing matrices correspond to the global fits reported in \cite{ParticleDataGroup:2024cfk}.
$\theta_t$ refers to the critical exponent along the direction of the corresponding coupling in the top-dominated regime (calculated at $k = 10^{1300}\,\rm GeV$), while $\theta_{UV}$ refers to the same critical exponents at the UV fixed-point at $10^{26,000}\,\rm GeV$. The critical exponents in the CKM elements refer to eigen-directions that are linear combinations of the matrix elements.}
\label{tab:UV-IRdata}
\endgroup
\end{table*}

\section{Matching prescriptions}\label{app:Dmatching}
The values of the Yukawa couplings and the Higgs quartic coupling can be converted into pole masses with the help of matching prescriptions.

For the top Yukawa coupling, we use the 2-loop matching prescription from \cite{Buttazzo:2013uya}. For the other fermions, we use the matching prescription from \cite{Huang:2020hdv}.

For the Higgs quartic coupling, our matching prescription works as follows: we know that within high-precision studies of RG flows in the SM starting at the Planck scale, the stability bound, $\lambda_H(k = M_{\rm Planck})=0$, translates into a Higgs mass of $m_H \approx 125 \, \rm GeV$ at a top pole mass of $m_{\rm top} \approx 171\, \rm GeV$, cf.~\cite{Bezrukov:2012sa,Buttazzo:2013uya}. Accordingly, we work with an $f_y$ such that $M_t \approx 171 \, \rm GeV$ is a consequence of using the 2-loop matching prescription for the top Yukawa coupling, cf.~\cite{Buttazzo:2013uya}, and adjust $f_{\lambda}$ such that we achieve $\lambda_H(k=M_{\rm Planck})\approx0$.

Going beyond our leading-order approximation to the beta functions, the same UV values for the couplings as reported in Tab.~\ref{tab:UV-IRdata}, will result in (slightly) different predictions for the IR values. We emphasize that this difference can be compensated for all Yukawa couplings by adjusting $f_y$ as well as the UV values of 11 of the 12 Yukawa couplings, namely all those that are asymptotically free at the UV fixed point and can therefore be chosen within a range of values at $k_{\rm UV} = 10^{25,000}\, \rm GeV$.

\section{Scaling regimes for the CKM matrix elements}

\begin{figure}[!t]
    \centering
    \includegraphics[width=0.46\linewidth]{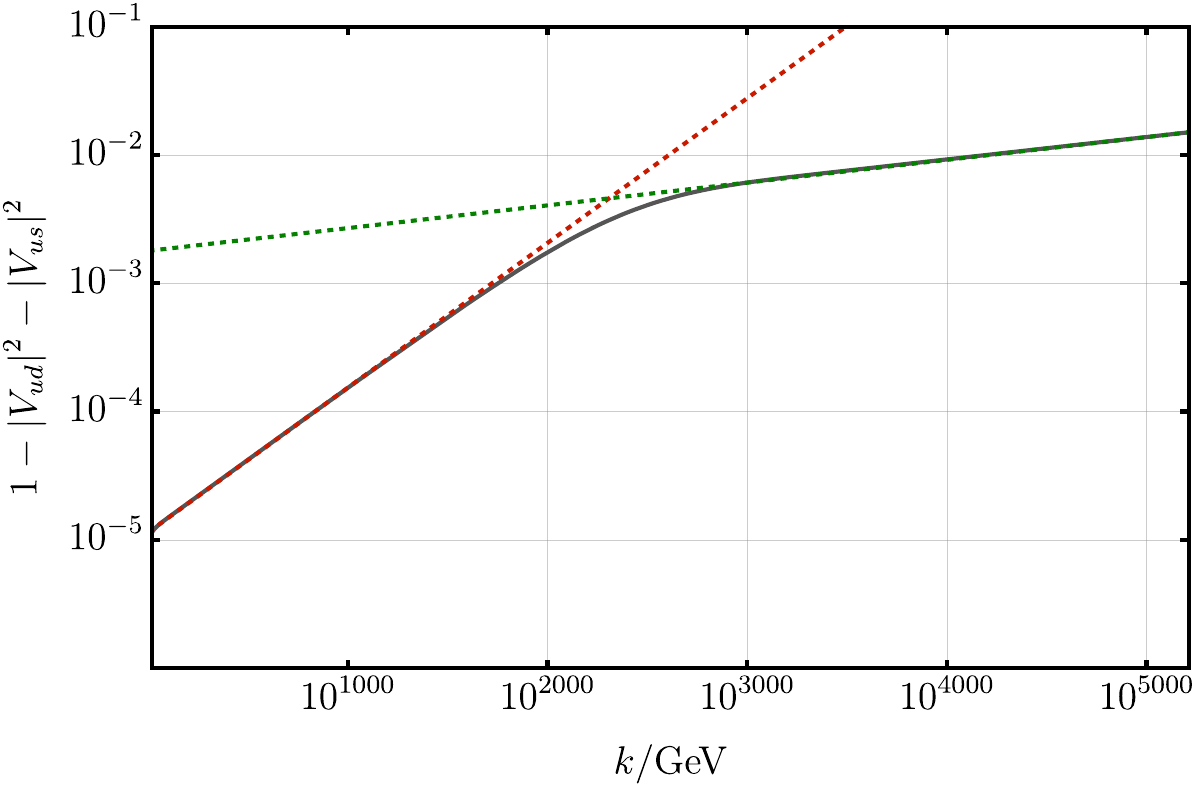}\quad
    \includegraphics[width =0.46\linewidth]{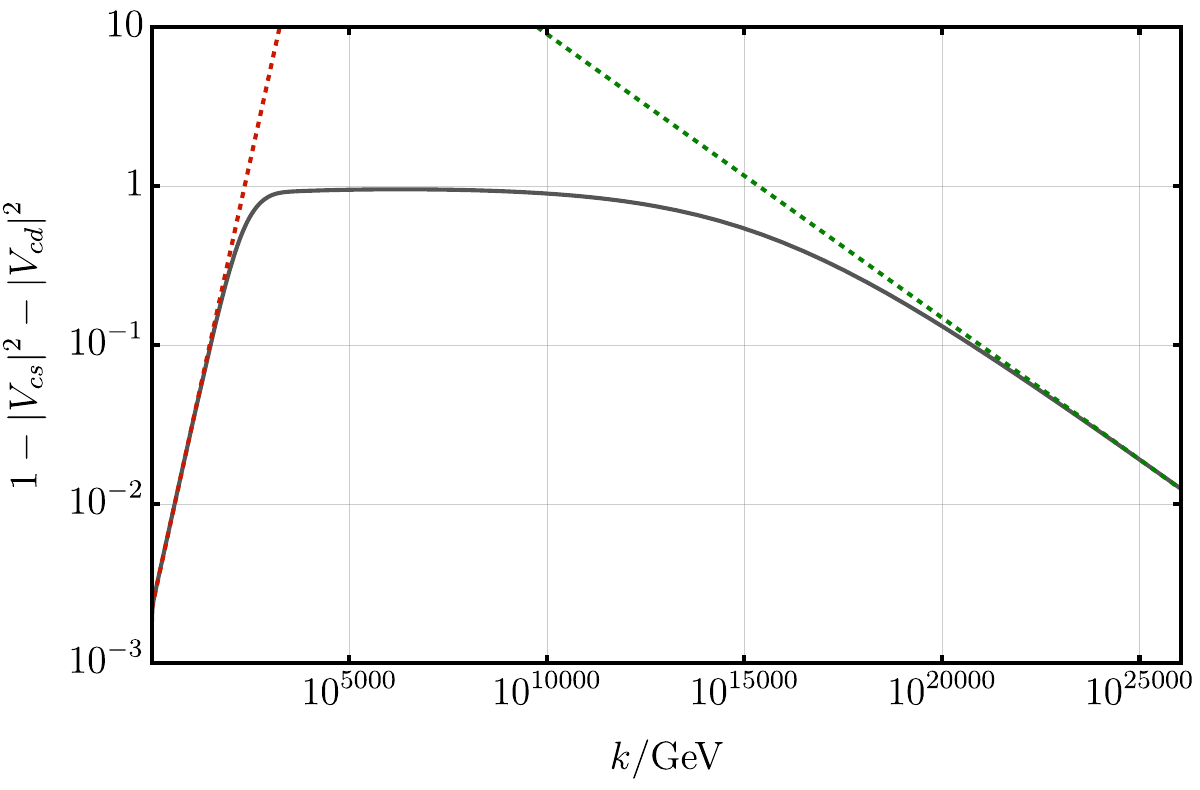}
    \caption{We show the RG flow of the relation $X+Y= 1$ (left panel) and $Z+W=1$ (right panel). Both, $X+Y-1$ and $Z+W-1$, scale towards zero in the top-dominated regime, with a rate captured by the critical exponent $\theta_{\rm top-dominated}= - \frac{3}{16\pi^2}y_t^2 \approx 1.1\cdot 10^{-3} $ (cf.~dark red lines). However, only $X+Y-1$ already scales towards zero in the bottom-dominated regime, with a critical exponent $\theta_{\rm bottom-dominated}= - \frac{3}{16\pi^2}y_b^2= -1.8\cdot 10^{-4}$ (cf.~dark green line).}
    \label{fig:scalingplot}
\end{figure}

\begin{table}[!t]
\centering
\begingroup
\setlength{\tabcolsep}{4pt} % Default value: 6pt
\renewcommand{\arraystretch}{1.4} % Default value: 1
\begin{tabular}{|c|c|c|}
fixed line & regime & critical exponents \\  \hline\hline 
$X = 0 = Y $  & $\frac{y_{b,\star}^2}{16 \pi^2} =\frac{2}{9} f_y + \frac{5}{369} f_g$ & \\ 
$ W = 1 - Z$ & $k \gg M_{\rm trans}$ & $\left(\frac{3 y_b^2}{16 \pi^2},\frac{3 y_b^2}{16 \pi^2},\frac{3 y_b^2}{16 \pi^2},0\right)$ \\
& (deep UV) & \\ \hline \hline 
$W = 0 = Z $  & $\frac{y_{b,\star}^2}{16 \pi^2} =\frac{2}{9} f_y + \frac{5}{369} f_g$ & \\ 
$ X = 1 - Y$ & $k \gtrapprox M_{\rm trans}$ & $\left( - \frac{3 y_b^2}{16\pi^2},\frac{3 y_b^2}{16\pi^2},0,0\right)$   \\ 
& (UV) & \\ \hline \hline
$ X = 1- Y$& $\frac{y_{t,\star}^2}{16\pi^2} =\frac{2}{9}f_y + \frac{17}{369} f_g$ & \\ 
$W = 1 - Z$ & $k < M_{\rm trans}$ & $\left(-\frac{3 y_t^2}{16 \pi^2},-\frac{3 y_t^2}{16 \pi^2},-\frac{3 y_t^2}{16 \pi^2},0\right)$\\
$W  = X$ & (IR) & \\ \hline \hline
\end{tabular}
\caption{We present the fixed line configurations and their critical exponents relevant for the flow of the CKM matrix. The fixed line in the deep UV is has 3 relevant directions and one marginal corresponding to our UV completion. The intermediate line $X = 1- Y$, $W = 0 =Z$ has only one IR attractive direction, while the IR configuration has 3 irrelevant directions leading to the high accuracy of the relations $X = 1 -Y$, $W = 1- Z$, $W = Z$ in the IR, in agreement with observations.}
\label{tab:fixedlines}
\endgroup
\end{table}

The relation $X+Y=1$ is IR attractive throughout the top- dominated regime and also in the bottom-dominated regime, once the RG flow has departed from the deep UV fixed point. In contrast, $Z+W=1$ is only IR attractive throughout the top-dominated regime.  This can be understood from the number of irrelevant and relevant directions of the fixed-line configurations of the CKM elements in Table ~\eqref{tab:fixedlines}. In Fig.~\ref{fig:scalingplot}, we show the differences $X+Y-1$ and $Z+W-1$, alongside the expected scalings {$X+Y-1 \sim {\rm exp}(- \theta_i\, t)$}, where $i = {\rm top/bottom-dominated}$; and $Z+W-1 \sim {\rm exp}(- \theta_{\rm top-dominated}\, t)$. A comparison of the upper and lower panel also highlights that $Z+W=1$ is not IR-attractive in the bottom-dominated regime, resulting in our expectation that $X+Y=1$ is approached more closely than $Z+W=1$ in the IR. The experimental data conforms to this expectation.

\section{Neutrino mass bound}
\begin{figure}[]
\begin{center}
\includegraphics[width=0.5\linewidth]{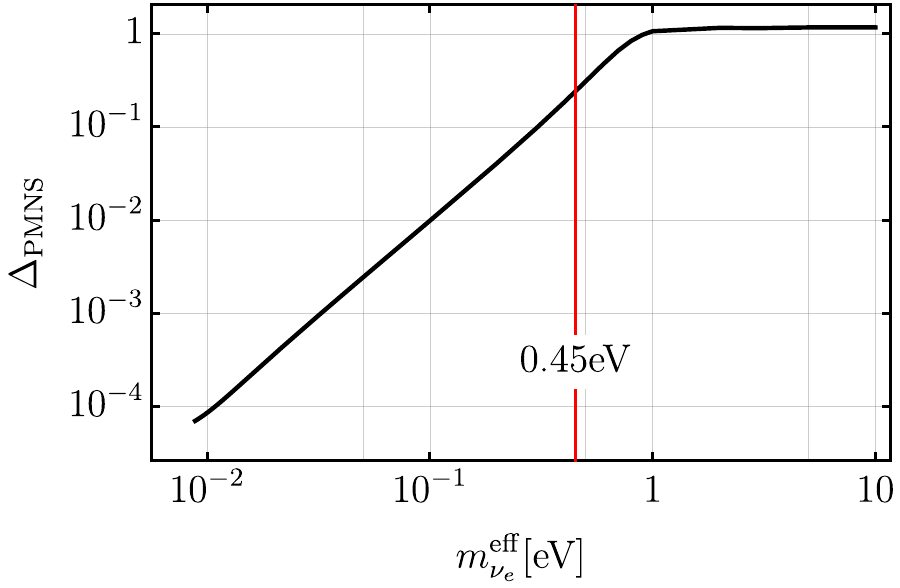}
\end{center}
\caption{\label{fig:neutrino-mass-bound} 
 We show the relative change in PMNS matrix elements under the RG flow as a function of the effective electron neutrino mass $m_{\nu_e}^{\rm eff}$. The red line at $0.45 \;\rm eV$ corresponds to the latest upper bound on $m_{\nu_e}^{\rm eff}$ by the KATRIN experiment \cite{KATRIN:2024cdt}. On the right side of the red line we have the experimentally excluded region which leads to $\mathcal{O}(1)$ variations of the PMNS matrix along the scales $(171\;\rm GeV, 10^{10000}\;\rm GeV)$.
}
\end{figure}

We compute the change in the matrix elements of $|U|^2_{\rm PMNS}$ under variations of the effective electron neutrino mass $m_{\nu_e}^{\rm (eff)}$, cf.~Fig.~\ref{fig:neutrino-mass-bound}.  We quantify the variation of the PMNS matrix elements along their RG flow  as:
\begin{eqnarray}
    \Delta_{\rm PMNS} = \sqrt{\sum_{\{ij\}} \left(\Delta U_{ij}\right)^2} \;\;,\;\;
\end{eqnarray}
where $\Delta U_{ij}$ is the difference between the IR and UV value of the corresponding PMNS element. We read off the IR value at $k \approx 171 \;\rm GeV$ and the  UV value at $k \approx 10^{10000} \;\rm GeV$. The sum over the index pairs $(ij)$ is understood as a sum over the independent components of the PMNS matrix, i.e. $\{ij\} = \{e1 \,,\, e2 \,,\,\mu1\,, \mu2\}$.

Fig.~\eqref{fig:neutrino-mass-bound} shows that the relative change in the matrix elements is $\mathcal{O}(1)$ for effective electron neutrino masses of order eV. Towards lower values of the effective electron neutrino mass, there is a power-law decrease of $\Delta_{\rm PMNS}$.
In fact, below the latest upper bound $m_{\nu_e}^{\rm eff}$ by KATRIN \cite{KATRIN:2024cdt},  $\Delta_{\rm PMNS} \lesssim 0.1$. 
Thus, $\mathcal{O}(1)$ variations of the PMNS matrix are already excluded since they require $m_{\nu_e}^{\rm eff} > 1 \;\rm eV$.

Fig.~\ref{fig:neutrino-mass-bound} supports our finding that the measured values of the PMNS elements are far from a near-diagonal configuration, because they do not change under the RG flow, and can therefore escape being attracted towards a near-diagonal matrix. 

\end{appendix}

% Use your bibtex library, formatted by the SciPost style file.
\bibliography{references.bib}

%%%%%%%%%% END TODO: BIBLIOGRAPHY

\end{document}